\documentclass[nonacm,sigconf]{acmart}

\AtBeginDocument{%
  }

\copyrightyear{2025}
\acmYear{2025}
\setcopyright{rightsretained}
\acmConference[WWW Companion '25]{Companion Proceedings of the ACM Web Conference 2025}{April 28-May 2, 2025}{Sydney, NSW, Australia}
\acmBooktitle{Companion Proceedings of the ACM Web Conference 2025 (WWW Companion '25), April 28-May 2, 2025, Sydney, NSW, Australia}
\acmDOI{10.1145/3701716.3715469}
\acmISBN{979-8-4007-1331-6/2025/04}

\usepackage{enumitem}
\usepackage{soul}
\usepackage{balance}

\author{Hoang Cuong Nguyen}
\affiliation{%
  \institution{Swinburne University of Technology}
  \city{Melbourne}
  \country{Australia}
}

\author{Shahroz Tariq}
\affiliation{%
  \institution{CSIRO's Data61}
  \city{Sydney}
  \country{Australia}
}
\authornote{Corresponding Author. Email: \texttt{shahroz.tariq@data61.csiro.au}}

\author{Mohan Baruwal Chhetri}
\affiliation{%
  \institution{CSIRO's Data61}
  \city{Melbourne}
  \country{Australia}
}

\author{Bao Quoc Vo}
\affiliation{%
  \institution{Swinburne University of Technology}
  \city{Melbourne}
  \country{Australia}
}

\begin{document}

\title[Towards Effective Identification of Attack Techniques in Cyber Threat Intelligence Reports using LLMs]{Towards Effective Identification of Attack Techniques in Cyber Threat Intelligence Reports using Large Language Models}{\thanks{Accepted for publication at The Web Conference 2025 (WWW '25).}}

\begin{abstract}
This work evaluates the performance of Cyber Threat Intelligence (CTI) extraction methods in identifying attack techniques from threat reports available on the web using the MITRE ATT\&CK framework. We analyse four configurations utilising state-of-the-art tools, including the Threat Report ATT\&CK Mapper (TRAM) and open-source Large Language Models (LLMs) such as Llama2. Our findings reveal significant challenges, including class imbalance, overfitting, and domain-specific complexity, which impede accurate technique extraction. To mitigate these issues, we propose a novel two-step pipeline: first, an LLM summarises the reports, and second, a retrained SciBERT model processes a rebalanced dataset augmented with LLM-generated data. This approach achieves an improvement in F1-scores compared to baseline models, with several attack techniques surpassing an F1-score of 0.90. Our contributions enhance the efficiency of web-based CTI systems and support collaborative cybersecurity operations in an interconnected digital landscape, paving the way for future research on integrating human-AI collaboration platforms.

\end{abstract}





\maketitle

\section{Introduction} \label{sec:intro}
In today's rapidly evolving digital landscape, cybersecurity has emerged as a critical concern for organisations worldwide. Security Operations Centres (SOCs) play a pivotal role in defending against the increasing sophistication of cyber threats by leveraging advanced technologies, such as artificial intelligence and machine learning~\cite{jalalvand2024alert}. These technologies enhance the capacity to detect, analyse, and respond to threats in real time, thereby improving the resilience of digital infrastructures. Moreover, the integration of human-AI teaming and collaboration is gaining traction as a strategy to enhance the efficiency and effectiveness in different domains~\cite{ParisReeson2024,schleiger2024collaborative,irons2024towards} including cybersecurity operations\cite{baruwal2024towards,chhetri2024internet,tariq2024a2c,Tariq2025lbw}. By combining the analytical strengths of human experts with the rapid processing capabilities of AI, SOCs can more effectively manage the vast and complex data streams they encounter daily. Cybersecurity analysts often rely on Cyber Threat Intelligence (CTI) reports to remain informed about the ever-evolving threat landscape.

CTI reports are comprehensive documents that provide valuable insights into current and emerging cyber threats faced by organisations.  Typically produced by cybersecurity analysts or specialised agencies, these reports aid businesses and government entities in understanding the threat landscape and adopting proactive measures to safeguard their digital assets~\cite{rahman2023attackers}. Key components of a CTI report include (i) detailed descriptions of various cyber threats, such as malware, ransomware, phishing attacks, and advanced persistent threats (APTs); (ii) profiles of threat actors, including their motives, tactics, techniques, and procedures (TTPs); (iii) Indicators of Compromise (IOCs) like IP addresses, malware hashes, or domain names that signal potential breaches; and (iv) recommended mitigation strategies to counteract these threats.

However, the manual analysis of CTI reports poses significant challenges due to their often unstructured and verbose nature. Such reports can extend over dozens of pages, making it arduous for SOC analysts to swiftly extract critical information~\cite{tang2022advanced}. This inefficiency contributes to the broader issue of alert fatigue, with studies indicating that up to 70\% of SOC analysts feel overwhelmed by the volume of alerts, leading 43\% to disable alerts as a coping mechanism~\cite{trendmicro2021toolsprawl,Tariq2025}. Given that modern cybersecurity operations depend heavily on real-time, web-based collaboration and decision-making, addressing these inefficiencies is vital for sustaining effective threat defence.

To alleviate these challenges, automated CTI extraction methods have been developed, facilitating the identification of IOCs and TTPs from extensive web-sourced reports~\cite{li2022attackg, tram_ctid}. Despite advancements in AI and Natural Language Processing (NLP), several hurdles persist in automating CTI analysis: \textbf{(i) Domain Complexity}: CTI reports often contain specialised terminology distinct from standard English, hindering accurate extraction by generic NLP tools; \textbf{(ii) Verbosity}: The relevant information about cyber-attacks is often buried within lengthy documents. For instance, a 42-page report~\cite{clearsky2016dustysky} may only dedicate a few paragraphs to the actual attack details; and \textbf{(iii) Relationship Extraction}: Accurately capturing the relationships between entities, such as attackers, tools, and victims, is essential for understanding TTPs, yet current NLP systems struggle with this intricate task~\cite{mu2018understanding}.

This study aims to address these challenges by exploring innovative approaches that enhance the automated extraction and utilisation of CTI reports, ultimately empowering SOCs to make more informed and timely decisions in the face of evolving cyber threats. Given these challenges, this research explores the following research questions:
\begin{itemize}
    \item  \textbf{RQ1:} How effective are standalone vanilla large language models (LLMs) in CTI extraction?
    \item \textbf{RQ2:} Can LLM-based augmentation improve the performance of automated CTI extraction methods?
\end{itemize}

\begin{figure*}[t]
    \centering
    \includegraphics[trim={17pt 16pt 17pt 17pt},clip,width=1\linewidth]{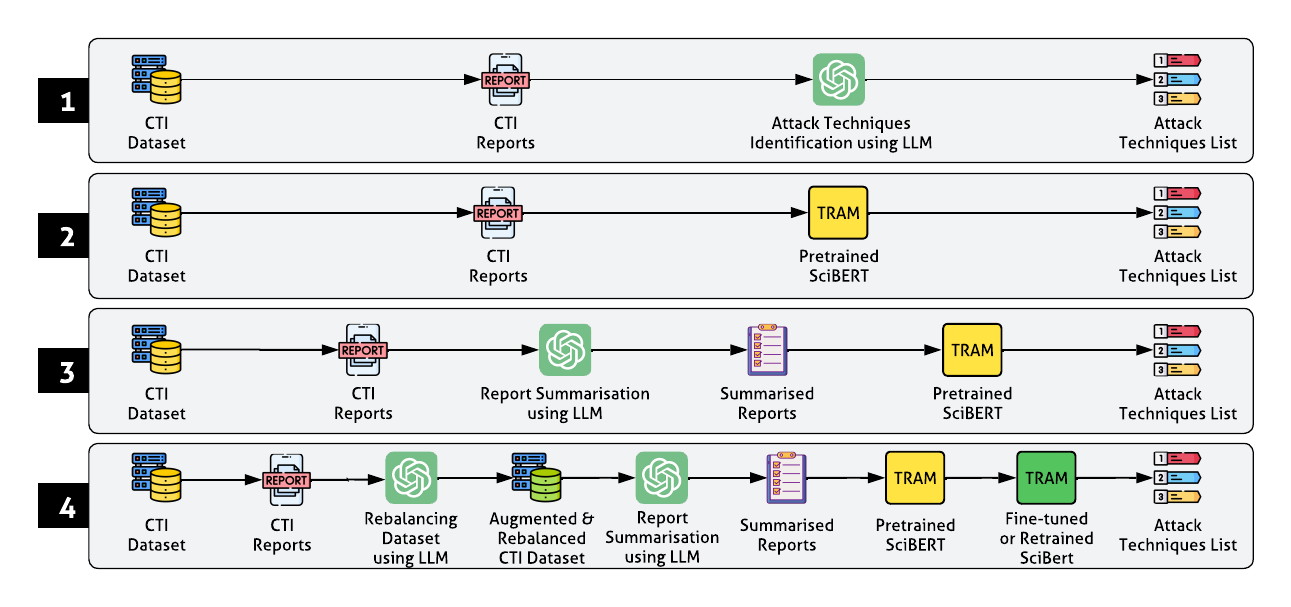}
    \caption{Our evaluation methodology uses four configurations.}
    \label{fig:pipeline}
\end{figure*}

In answering these research questions, we make the following contributions:
\begin{enumerate}
    \item\textbf{Comprehensive Evaluation}: We evaluate the CTI extraction method using four configurations, as shown in Figure~\ref{fig:pipeline}. Our evaluation highlights the strengths and weaknesses of each configuration, with the best-performing baseline (i.e., TRAM with original SciBert), achieving an F1-score of just over 0.4 due to overfitting and class imbalances.
    \item \textbf{Novel Extraction Pipeline}: Inspired by recent developments in LLMs, we propose a two-step pipeline: \textit{(i) CTI Report Summarisation using GPT-3.5}: Reducing report verbosity to focus on key threat information; and \textit{(ii) Retrained SciBERT Model}: Using a SciBERT model~\cite{beltagy2019scibert} trained on a rebalanced dataset to address class imbalance and improve classification accuracy, resulting in an F1-score of over 0.90 in identification of several attack techniques.
\end{enumerate}

The remainder of this paper is organised as follows: \autoref{sec:related} reviews the current progress and limitations in CTI extraction. \autoref{sec:method} details the evaluation methodology and proposed approach. \autoref{sec:results} presents the evaluation results. \autoref{sec:conc} concludes this work and outlines the limitations and future research directions\footnote{Our code is available here: \url{https://github.com/hoangcuongnguyen2001/SciBERT-for-Technique-Classification}}.

\section{Background and Related Works} \label{sec:related}
This section provides an overview of CTI, its extraction and sharing processes, and the current limitations of existing CTI extraction methods, highlighting the need for improved methodologies.
\subsection{Cyber Threat Intelligence}
Cyber Threat Intelligence (CTI) serves as a preventative defence mechanism against cyber-attacks. According to the National Institute of Standards and Technology (NIST), CTI is defined as ``threat information that has been aggregated, transformed, analysed, interpreted, or enriched to provide the necessary context for decision-making processes''~\cite{nist_threat_intelligence}.

The primary criterion for evaluating CTI is its actionability. According to Pawlinski et al.~\cite{pawlinski2014actionable}, actionable CTI must possess the following characteristics: \textit{(i) Relevance}: Applicable to the area of responsibility for recipients; \textit{(ii) Timeliness}: Information must be recent enough to be effective; delays can render CTI obsolete; 
\textit{(iii) Accuracy}: Information should be verified and error-free; 
\textit{(iv) Completeness}: Sufficient context to understand past cyber-attacks; and \textit{(v) Ingestibility}: Shared in a format that can be processed by recipient systems.

\subsection{CTI Extraction and Sharing}
Modern CTI extraction methods typically follow a standardised pipeline comprising the following steps: \textit{(i) Identifying Sources}: Selecting relevant threat reports for analysis; \textit{(ii) Report Crawling}: Automatically gathering reports from various repositories; \textit{(iii) Text Processing and Labelling}: Extracting and annotating relevant entities, such as Indicators of Compromise (IOCs) and Tactics, Techniques, and Procedures (TTPs); \textit{(iv) Text Summarisation}: Reducing verbosity while retaining key information using learning-based approaches; and \textit{(v) Processing Outputs}: Converting extracted data into formats suitable for downstream applications, such as knowledge graphs.
Methods such as AttackG~\cite{li2022attackg} and TRAM~\cite{tram_ctid} have been developed to automate CTI extraction. These methods leverage NLP techniques to identify attack techniques based on the MITRE ATT\&CK framework.

\noindent
\textbf{\textsc{Current Limitations of CTI Extraction Methods}. } Despite advancements, CTI extraction methods face significant challenges that hinder their effectiveness: \textit{(i)~Domain Complexity}: CTI reports contain cybersecurity-specific terminology that differs from standard English, making it difficult for general NLP models to process accurately. \textit{(ii)~Verbosity}: Many CTI reports are lengthy, with only a small portion containing actionable threat information. For example, a 42-page report may contain only a few sentences detailing the actual attack~\cite{clearsky2016dustysky}. \textit{(iii)~Relationship Extraction}: Extracting relationships between entities (e.g., attackers, tools, victims) is essential but remains challenging for existing NLP systems~\cite{mu2018understanding}. \textit{(iv)~Class Imbalance}: Many extraction methods suffer from class imbalance, where certain techniques are overrepresented while others are underrepresented. \textit{(v)~Replication Inconsistency}: Performance claims in research papers often fail to replicate on different datasets due to varying conditions and datasets.

These challenges contribute to low precision, recall, and F1-scores, highlighting the need for more effective CTI extraction methodologies. Our work attempts to address some of the limitations of existing CTI extraction methods through a comprehensive evaluation of multiple CTI extraction methods and the introduction of a novel extraction pipeline.

\section{Methodology}
\label{sec:method}
Our evaluation methodology consists of five key components: extraction method selection, dataset preparation, experimental design, and evaluation metrics.

\subsection{Extraction Method Selection}
To answer \textbf{RQ1}, we selected three variants of vanilla Llama2~\cite{touvron2023llama}, i.e., 7B, 13B, and 70B, under zero-shot prompting. 
To answer \textbf{RQ2}, we selected TRAM, a SciBERT-based method designed to classify sentences into the 50 most prevalent techniques within the MITRE ATT\&CK framework~\cite{tram_ctid}, as a base model and used different configurations with and without the LLMs-based augmentations for evaluation.

\subsection{Ground Truth Datasets}
Two annotated datasets were used as ground truth for evaluation:

\begin{enumerate}
    \item \textbf{\textit{Adversary Emulation Library (AEL)}}: This dataset comprises concise reports on attack campaigns (e.g., APT29, Carbanak, FIN6) annotated with MITRE ATT\&CK technique IDs~\cite{adversary_emulation_library}.
    \item \textbf{\textit{Attack-Technique-Dataset (ATD)}}: This dataset contains longer reports (e.g., OceanLotus, Sowbug, MuddyWater) annotated with detailed technique information~\cite{attack_technique_dataset}.
\end{enumerate}

Reports were preprocessed to remove technique IDs, hyperlinks, and extraneous content to ensure unbiased evaluation. Techniques outside the top 50 most prevalent techniques in the MITRE ATT\&CK framework were excluded to align with TRAM's training data.

\subsection{Experimental Design}
As shown in Figure~\ref{fig:pipeline}, we use four experimental settings to assess and compare the CTI extraction methodologies:
\begin{enumerate}
    \item \textbf{\textit{Standalone LLM for CTI Extraction}}: The zero-shot prompting capabilities of open-source LLMs (Llama2) were evaluated.  
    Precision, recall, and F1-score were computed, along with counts of true positives, false positives, and false negatives.

    \item \textbf{\textit{Original TRAM Configuration}}:  
    TRAM's performance was evaluated using a pre-trained SciBERT model with confidence thresholds of 25\% and 80\%. We will denote it as Original SciBERT.
   
   \item \textbf{\textit{TRAM with LLM-based Summarisation}}: This configuration uses summarised CTI reports generated by an LLM (GPT-3.5), followed by SciBERT classification at confidence levels of 25\% and 75\%. We will denote it as aCTIon. \textit{Note: This selection is based on best performance settings.}
    
    \item \textbf{\textit{TRAM with LLM-based Summarisation, Rebalancing and Retraining}}: In this configuration, we are Augmenting underrepresented techniques using GPT-3.5 and downsampling overrepresented techniques.  CTI reports were summarised using GPT-3.5 to reduce verbosity and retain relevant content related to attack techniques. Then, summarised reports were processed by a SciBERT model retrained on the rebalanced dataset. We used three settings for training i.e., retraining, fine-tuning and retraining with  5-fold cross-validation, resulting in a total of 9 configurations for evaluation of TRAM. 
\end{enumerate}

\noindent
\textbf{\textsc{Evaluation Metrics}. }
The performance of each extraction method was evaluated using standard classification metrics: precision, recall and F1-score. \textit{Note: We only report the F1-score in this work due to space constraints.} Additional analysis included counts of true positives, false positives, and false negatives. For AttackG and LLMs, results were considered correct if the extracted technique names or IDs matched the ground truth. For TRAM, exact matches of both technique names and IDs were required.

\begin{table}[t]
    \centering
    \caption{Comparison of false positives and false negatives between Llama2 models for AEL dataset. The `-' sign represents a false negative and the `+' sign represents false positives.}
    \label{tab:attackg_llm_fpfn}
    \begin{tabular}{l|c|c|c}
        \hline
        \textbf{Report}  & \textbf{Llama2-7B} & \textbf{Llama2-13B} & \textbf{Llama2-70B} \\
        \hline
        APT29     & -2 / +9  & -2 / +10 & 0 / +9 \\
        Carbanak  & -1 / +5  & -5 / +8  & -3 / +8 \\
        FIN6      & -21 / +7 & -24 / +8 & -22 / +19 \\
        FIN7      & -4 / +7  & -3 / +14 & -2 / +7 \\
        menuPass  & -4 / +5  & -4 / +11 & -5 / +12 \\
        OilRig    & -1 / +5  & -2 / +5  & -2 / +8 \\
        \hline
    \end{tabular}
\end{table}
\begin{table}[t]
    \centering
    \caption{True positives for Llama2 models.}
    \label{tab:attackg_llm_tp}
    \begin{tabular}{l|c|c|c|c}
        \hline
        \textbf{Report} & \begin{tabular}[c]{@{}c@{}}\textbf{Ground}\\\textbf{Truth}\end{tabular} & \begin{tabular}[c]{@{}c@{}}\textbf{Llama2}\\\textbf{7B}\end{tabular} & \begin{tabular}[c]{@{}c@{}}\textbf{Llama2}\\\textbf{13B}\end{tabular} & \begin{tabular}[c]{@{}c@{}}\textbf{Llama2}\\\textbf{70B}\end{tabular} \\ 
\hline
        APT29    & 3 & 1 & 0 & 3 \\
        Carbanak & 6 &  5 & 1 & 3 \\
        FIN6     & 24 &  3 & 0 & 2 \\
        FIN7     & 4 &  0 & 1 & 2 \\
        menuPass & 5 &  1 & 1 & 0 \\
        OilRig   & 4 &  3 & 2 & 2 \\
        \hline
    \end{tabular}
    
\end{table}
\begin{table}[t]
    \centering
    \caption{Average performance of Llama2 models.}
    \label{tab:attackg_llm_metrics}
    \begin{tabular}{l|c|c|c}
        \hline
        \textbf{Method} & \textbf{Precision} & \textbf{Recall} & \textbf{F1-Score} \\
        \hline
        Llama2-7B      & \textbf{0.2403}   & 0.3736   & \textbf{0.2733}   \\
        Llama2-13B     & 0.0911   & 0.2792   & 0.1199   \\
        Llama2-70B     & 0.1733   & \textbf{0.4306}   & 0.2384   \\
        \hline
    \end{tabular}
\end{table}

\section{Results}
\label{sec:results}
This section presents the evaluation results of standalone open-source LLMs for CTI extraction, followed by a comparative analysis of different TRAM configurations, highlighting the impact of model refinements and dataset rebalancing on performance.

\subsection{Evaluation of Standalone Open-Source LLMs}
To address RQ1, we evaluate the performance of LLama2 using six reports from the AEL dataset, each under 500 words. This ensured manageable processing times and mitigated dependency explosion issues. Table~\ref{tab:attackg_llm_fpfn} compares the number of false positives and false negatives for various versions of LLama2 (7B, 13B, and 70B). 
Notably, the larger 70B LLama2 model tended to produce more false positives in several instances. For example, in the case of FIN6, LLama2-7B, and LLama2-13B produced 7, and 8 false positives, respectively, whereas LLama2-70B produced 19 false positives, more than twice that of LLama2-13B. Future work should explore whether fine-tuning the LLama2 model could reduce the number of false positives and negatives.

Table~\ref{tab:attackg_llm_tp} displays the number of true positives detected by each model and the overall precision, recall, and F1-score for each method are summarised in Table~\ref{tab:attackg_llm_metrics}. All models performed poorly, detecting only a fraction of the true positives in many scenarios, as evident in Tables~\ref{tab:attackg_llm_tp} and \ref{tab:attackg_llm_metrics}. Although the LLama2 70B model detected the most true positives, the overall performance of LLama2 7B was slightly better due to the higher number of false positives generated by LLama2 70B. This low performance highlights the complexity of the task and the need for specialised methods for CTI extraction. Our next research question aims to explore this very direction.

\begin{figure}[t]
    \centering
    \includegraphics[width=1\linewidth]{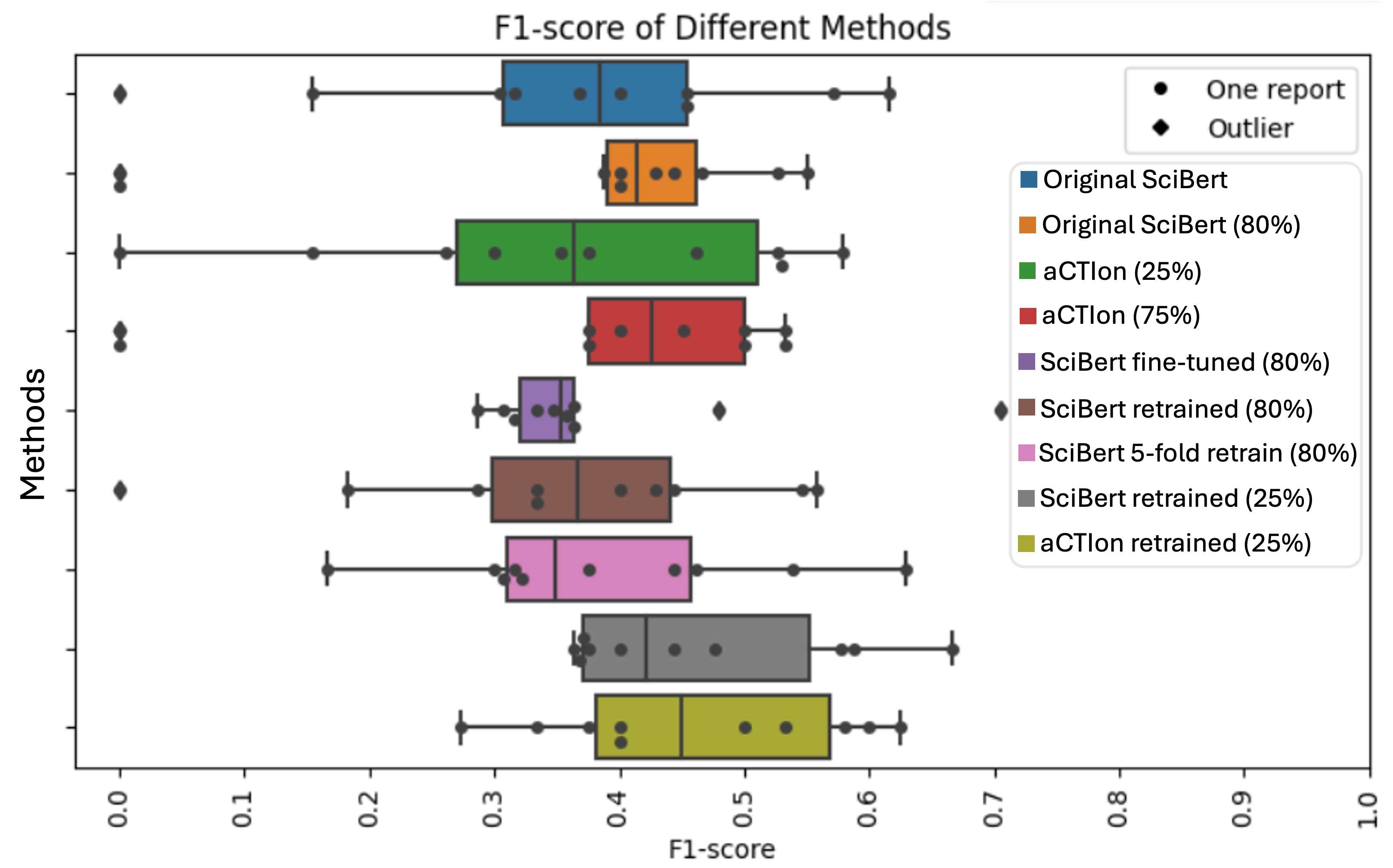}
    \caption{Performance of TRAM using different Configurations proposed in our work on ATD dataset.}
    \label{fig:tram_f1_improv}
\end{figure}

\begin{table}
\centering
\caption{Classification results for selected techniques using the retrained SciBERT model on the ATD dataset.}
\label{tab:sci_bert_results}

\begin{tabular}{l|c|c|c|c} 
\hline
\begin{tabular}[c]{@{}l@{}}\textbf{Technique }\\\textbf{ID}\end{tabular} & \textbf{Precision} & \textbf{Recall} & \textbf{F1-Score} & \begin{tabular}[c]{@{}c@{}}\textbf{Sentences in }\\\textbf{Test Set}\end{tabular} \\ 
\hline
T1056.001 & 0.8000 & 0.8649 & 0.8312 & 37 \\
T1057 & 0.7931 & \textbf{1.0000} & 0.8846 & 46 \\
T1059.003 & 0.6848 & 0.7975 & 0.7368 & 79 \\
T1070.004 & 0.9000 & 0.9474 & \textbf{0.9231} & 76 \\
T1566.001 & 0.8679 & 0.9583 & 0.9109 & 48 \\
T1570 & \textbf{1.0000} & 0.3125 & 0.4762 & 16 \\
\hline
\end{tabular}

\end{table}

\subsection{Comparison of TRAM Configurations}

To address RQ2, we evaluated TRAM under different configurations and confidence levels (25\% and 80\%). The results, shown in Figure~\ref{fig:tram_f1_improv}, highlight the impact of configuration changes. We can observe that first summarising the report with GPT-3.5 and then SciBERT classification results in slightly better performance than the default setting of SciBERT. 

To address class imbalance and overfitting in TRAM, we retrained the SciBERT model on a rebalanced dataset. This approach resulted in a median F1-score increase of approximately seven percentage points compared to the baseline model, as shown in Figure~\ref{fig:tram_f1_improv}. Classification results for a few selected techniques using the best-performing retrained SciBERT model are presented in Table~\ref{tab:sci_bert_results}. We observed that this model performs well on many of the top 50 most prevalent techniques in the MITRE ATT\&CK framework, achieving an F1-score of up to 0.92.

\section{Conclusion}
\label{sec:conc}
This work evaluated state-of-the-art Cyber Threat Intelligence extraction methods, highlighting key challenges such as class imbalance, overfitting, and the complexity of cybersecurity texts. To address these issues, we proposed a novel pipeline that combines GPT-3.5 for report summarisation and a retrained SciBERT model to improve classification accuracy using rebalanced data augmented by an LLM. This approach resulted in a seven-percentage-point increase in the F1-score compared to the baseline model and achieved above 0.90 F1-score for several attack techniques. Despite these improvements, challenges remain in classifying underrepresented techniques and reducing false positives. Future work should focus on integrating human-AI collaboration to enhance extraction accuracy and exploring fine-tuned LLMs for more effective CTI analysis. These contributions lay the groundwork for more reliable automated CTI systems to support cybersecurity operations.
\begin{acks}
This work was supported by Commonwealth Scientific and Industrial Research Organisation (CSIRO) Collaborative Intelligence (CINTEL) Future Science Platform. The diagram have been designed using images from flaticon.com
\end{acks}

\bibliographystyle{ACM-Reference-Format}
\balance
\bibliography{0_references.bib}

\end{document}